\documentclass[12pt]{article}
\usepackage{amsmath}
\usepackage{amssymb}
\usepackage{amscd}
\usepackage{graphics}
\usepackage{graphicx} 
\usepackage{epsfig}      
\usepackage{color}            
\usepackage{epic}             
\usepackage{eepic}            
\usepackage{rotating}         
\usepackage{type1cm}  
\usepackage{appendix}
\usepackage{graphicx}
\usepackage{verbatim}
\usepackage{epsf}
\usepackage{latexsym}
\usepackage{amsmath,amsfonts,amssymb,amsthm} 

\def\bseq{\begin{subequation}}  
\def\eseq{\end{subequation}}
\def\bsea{\begin{subeqnarray}}  
\def\esea{\end{subeqnarray}}

\newcommand{\bbox}{\lower.2ex\hbox{$\Box$}}

\newcommand{\beq}{\begin{equation}}
\newcommand{\eeq}{\end{equation}}
\newcommand{\bea}{\begin{eqnarray}}
\newcommand{\eea}{\end{eqnarray}}
\newcommand{\ena}{\end{eqnarray}}

\newcommand{\Tr}{{\rm Tr}}
\renewcommand{\(}{\left(}
\renewcommand{\)}{\right)}
\renewcommand{\[}{\left[}
\renewcommand{\]}{\right]}

\newcommand{\be}{\begin{equation}}
\newcommand{\ee}{\end{equation}}

\numberwithin{equation}{section}

\begin{document}
\setcounter{page}{0}
\begin{titlepage}
\titlepage
\begin{flushright}
LPTENS-09/21\\
\end{flushright}
\vskip 4cm
\begin{center}
\LARGE{\Huge Spin chain for the deformed ABJM theory \\}
\LARGE{\Huge  }
\end{center}
\vskip 0.5cm \centerline{{\bf  Davide Forcella$^{a}$ \footnote{\tt forcella@lpt.ens.fr} \ \ \
    Waldemar Schulgin$^{a,b}$ \footnote{\tt waldemar.schulgin@lpt.ens.fr}}}
\medskip
\footnotesize{

\begin{center}\it{
$^a$ Laboratoire de Physique Th\'eorique de l'\'Ecole Normale Sup\'erieure \\
and CNRS UMR 8549\\
24 Rue Lhomond, Paris 75005, France\\
\medskip
$^b$ Laboratoire de Physique Th\'eorique et Hautes Energies,\\
UPMC Univ Paris 06, Boite 126, 4 place Jussieu,
F-75252 Paris Cedex 05 France}
\end{center}}

\bigskip

\begin{abstract}

In this short note we begin the analysis of deformed integrable Chern-Simons theories. We construct the two loop 
dilatation operator for the scalar sector of the ABJM theory with $k_1 \ne -k_2$ and we compute the anomalous dimension
of some operators.

\end{abstract}

\vfill
\begin{flushleft}
\end{flushleft}
\end{titlepage}

\newpage

\tableofcontents

\section{Introduction}

In the context of the $AdS_5/CFT_4$ correspondence a very interesting development was the understanding of the existence of an
integrable structure at the both sides of the correspondence \cite{Minahan:2002ve,Bena:2003wd}. In the $\mathcal{N}=4$ $SU(N)$ field theory the one loop dilatation operator in the scalar sector was identified with the Hamiltonian of an integrable spin chain \cite{Minahan:2002ve}. Many and very interesting developments followed, see for example \cite{Berenstein:2002jq}-\cite{Gromov:2009tv}. 
In this note we will be mainly interested in studying the integrability properties of the field theories with less supersymmetry.
In four dimensions to remain in the perturbative regime, which allows a field theory computation, one is  forced to take orbifold or marginal deformations of the original $\mathcal{N}=4$ theory \cite{Berenstein:2004ys}-\cite{Solovyov:2007pw}.

Recently we gained a better understanding of the $AdS_4/CFT_3$ correspondence \cite{Schwarz:2004yj}-\cite{Jafferis:2008qz}.
Indeed, it turns out that the 
three dimensional conformal field theories are Chern-Simons theories with matter.
In particular, the authors of \cite{Aharony:2008ug} proposed a field theory dual to the $\mathbb{C}^4/\mathbb{Z}_k$ singularities, the so called ABJM theory.
This is an $\mathcal{N}=6$ Chern-Simons matter theory with 
gauge group $U(N) \times U(N)$ and two Chern-Simons levels satisfying the  constraint $k_1 + k_2 = 0$.
This theory appears to be integrable at least at the leading order in perturbation theory. Namely, the two loop dilatation operator can be identified with the Hamiltonian of an integrable spin chain \cite{Minahan:2008hf}.
 Many nice developments followed also in this context, see for example \cite{Bak:2008cp}-\cite{Agarwal:2008pu}.

In particular we are interested to understand if integrability is present in the less supersymmetric theories.
One could think that the possible generalizations of the basic example in three dimensions, the ABJM theory,  are 
very similar to the related generalizations of the $\mathcal{N}=4$ four dimensional case but they are slightly different.

To compute the field theory dilatation operator, it is important that the theory has a weak coupling 
limit in which the elementary fields have canonical scaling dimensions. In four dimensions this is possible if the 
superpotential is a cubic function of the chiral superfields, while in three dimensions it is possible 
if the superpotential is a quartic function. This simple observation points out that in three dimensions there 
could be more theories which can be analyzed perturbatively than in four dimensions. Indeed, it turns out that 
in three dimensions also the non-orbifold theories can have a perturbative limit \cite{Gaiotto:2007qi,Gaiotto:2009mv,Jafferis:2008qz}. 

The second observation is due
to the presence of Chern-Simons levels that do not exist in the four dimensional case. 
There are  Chern-Simons levels $k_i$ associated to every gauge group. 
They are integer numbers and we can vary their values without spoiling the superconformal symmetry.
It turns out that, for a class of $\mathcal{N}=2$ Chern-Simons matter theories, if $\sum k_i = 0$ the field  theory moduli space has a four complex dimensional branch that is a Calabi-Yau cone and can be understood as the space transverse to the M2 brane \cite{Martelli:2008si,Hanany:2008cd}. If instead $\sum k_i \ne 0$ the four dimensional branch typically disappears and this effect can be interpreted as turning on a Roman's mass $F_0$ in the type IIA limit \cite{Gaiotto:2009mv,Gaiotto:2009yz}. Let us suppose that a theory has an integrable structure for some specific relations among the $k_i$ such that they satisfy $\sum k_i = 0$. It easy to see that there exist two possible  interesting deformations
 of this integrable point. We can move in the space of possible integer values of the $k_i$ in such a way that 
we preserve the constraint or in a way in which we break the constraint. It is important to underline that these kind of
deformations do not exist in four dimensions and  offer a new laboratory for studying integrability in the weak coupling regime. 

In this paper we start the analysis of these deformed theories. 
We take as basic example the ABJM theory and  deform it in such a way that $k_1 + k_2 \ne 0$. 
We plan to return to the other type of deformation in the near future. 
To be sure to remain in the perturbative regime it is important to deform the theory in such a way that it preserves at least $\mathcal{N}=3$ supersymmetry in three dimensions. Indeed, for $\mathcal{N}>2$ the Chern-Simons matter field theories are completely 
specified by the gauge group, the matter content, and the Chern-Simons levels, and they have a weak coupling limit for large values of $k_i$. These theories have a quartic superpotential and 
could be dual to the non-orbifold M theory backgrounds.

The organization of the paper is as follows. In Section 2 we introduce our main example. In Section 3 we rewrite the theory in the explicit invariant form under the global symmetries. In Section 4 we compute the two loop mixing operator for the scalar sector of the theory. In Section 5 we compute the anomalous dimension of some operators. We observe that the degeneracy which due to integrability is present in the ABJM theory is lifted in the generic $k_1 \ne - k_2$ case. We finish with some conclusions and the appendix collects some useful formulae which we used in the main text.




\section{The deformed ABJM}

We are interested in studying the Chern-Simons theories described by the following action
\begin{equation}\label{SN3} 
S= \frac{k_1}{4\pi} S_{CS}(V_{(1)}) + \frac{k_2}{4\pi} S_{CS}(V_{(2)}) + S_{kin}(Z^i,Z_{i}^{\dagger},W_j, W^{j \dagger}) + \int  \ d^{2} \theta W(Z^i,W_j) + c.c.  \ ,\nonumber
\end{equation}
where
\begin{eqnarray}\label{inter}
& & S_{CS}(V_{(l)})= \int d^3x  \ \Tr \[ \epsilon^{\mu \nu \lambda} \( A_{(l) \mu} \partial_{\nu} A_{(l)  \lambda} + \frac{2i}{3} A_{(l)  \mu} A_{(l)  \nu} A_{(l)  \lambda} +i \bar{\chi}_{(l)} \chi_{(l)} - 2 D_{(l)} \sigma_{(l)} \)\]\ ,  \nonumber \\
& & S_{kin}(Z^i,Z_{i}^{\dagger},W_j, W^{j \dagger}) = \int\  d^{4}\theta \  \Tr  \ \( Z_{i}^{\dagger}e^{-V_{(1)}} Z^{i}e^{V_{(2)}} +  W^{j \dagger}e^{-V_{(2)}} W_{j}e^{V_{(1)}} \)  \ , \nonumber \\
& & W(Z^i,W_j)=\frac{2\pi }{k_1} \Tr \left(Z^i W_i Z^j W_j\right)+\frac{2\pi }{k_2} \Tr \left(W_i Z^i W_j Z^j\right) \ .
\end{eqnarray}

\begin{figure}[ht!]
\begin{center}
  \includegraphics[scale=0.6]{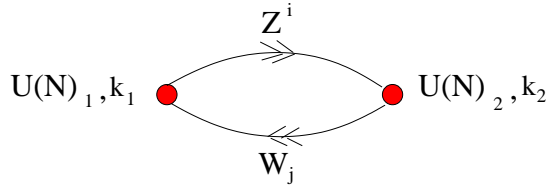}
\caption{\small The quivers for the ABJM theory with generic Chern-Simons levels.}
\label{quivk1k2}
\end{center}
\end{figure}

It is a three dimensional Chern-Simons theory with matter. The gauge group is $U(N)_1 \times U(N)_2$ and 
the $\mathcal{N}=2$ bifundamental chiral superfields $Z^i$ and $W_j$ transform  in the 
fundamental of the first factor of the gauge group and antifundamental of
the second one and vice versa for $Z^{\dagger}_i$ and $W^{\dagger j}$(see figure \ref{quivk1k2}). 
$k_1$, $k_2$ are integer numbers which  we 
call  from now on Chern-Simons levels. 
The three dimensional theory represented by the Lagrangian (\ref{SN3}) is $\mathcal{N}=3$ superconformal. It admits a perturbative limit for the large values of the $k_i$. The Lagrangian has $SU(2)_R \times SU(2)$ global symmetry, where the first factor is the R symmetry associated to the $\mathcal{N}=3$ superconformal symmetry, while the second $SU(2)$ is a global symmetry under which $Z^i$ and $W_j$ transform in the fundamental representation.

In the particular case $k_1= -k_2$ the supersymmetry of the Lagrangian is enhanced to  $\mathcal{N}=6$ and the global symmetry group to $SU(4)_R$. In this case the lower bosonic components\footnote{We use the same symbols $Z^i$, $W_j$
for the superfields and for their lowest scalar components. We hope this will not cause too much confusion.} of the chiral superfields can be organized in the fundamental representation ${\bf 4}$: $Y^A=(Z_1,Z_2,W^\dagger_1,W^\dagger_2)$ and the upper ones in the antifundamental. Indeed, in this limit the Lagrangian (\ref{SN3}) reduces to the ABJM one \cite{Aharony:2008ug} which is supposed to describe the three dimensional superconformal field theories living on N M2 branes at $\mathbb{C}^4/\mathbb{Z}_k$ singularities. In this particular case the theory is integrable in the planar limit at least at the two loop order.
The first check for the presence of integrability in the ABJM comes from the computation of the two loop mixing matrix of anomalous dimensions for the scalar sector. Due to integrability  the mixing matrix of anomalous dimensions is identical to an integrable Hamiltonian of the $SU(4)$ spin chain with the sites transforming  under  ${\bf 4}$ and ${\bf\bar{4}}$ \cite{Minahan:2008hf}.  

A natural question is  if the generic theory in eq.  (\ref{SN3}) is still integrable. In the case $k_1 \ne -k_2$ the supersymmetry and the global symmetries are  reduced, and the theory is supposed to be dual to some flux background. The four dimensional Calabi-Yau branch in the field theory moduli space disappears and the theory is proposed to be dual to a type IIA background with the Romans mass $F_0$ turned on: $k_1 + k_2= F_0$ \cite{Gaiotto:2009mv}. It is important to stress that this kind of deformation is not an orbifold deformation and this is a peculiarity of the Chern-Simons theories.

In this paper we would like to do the first step towards understanding the question concerning integrability for this theory. We compute the dilatation operator in the scalar sector at the leading order which we use then to find  anomalous dimensions of some operators. 
%
To make the computation more transparent we rewrite the eq. (\ref{SN3}) in such a way that the $SU(2)_R\times SU(2)$ symmetry becomes apparent. We  group the scalar fields into the tensors $O^a_i$ and  $O^{\dagger i}_a$, where the indices from the beginning of the alphabet correspond to the $SU(2)_R$ and from the middle to the $SU(2)$ symmetry group
\begin{equation}\label{OO}
O=\begin{pmatrix}      
Z^\dagger_1 & W_1  \\  
Z^\dagger_2 & W_2       
\end{pmatrix} \ ,\qquad  \qquad   
O^{\dagger}=\begin{pmatrix}
Z^1 & Z^2 \\
W^{\dagger 1} & W^{\dagger 2}
\end{pmatrix} \ .
\end{equation}
They transform in the ${\bf (2,2)}$ of $SU(2)_R \times SU(2)$ as $U O V^\dagger$, $V O^\dagger U^\dagger$, where $U \in SU(2)$ and $V \in SU(2)_R$.  

The class of the gauge invariant operators we are interested in has the form
\begin{equation}\label{op}
\mathcal{O}=\Tr \( O^{\dagger i_1}_{a_1} O^{a_2}_{i_2} O^{\dagger i_3}_{a_3} O^{a_4}_{i_4}.......O^{\dagger i_{2L-1}}_{a_{2L-1}} O^{a_{2L}}_{i_{2L}}\)\chi^{a_1i_2 a_3 i_4...a_{2L-1},i_{2L}}_{i_1 a_2 i_3 a_4...i_{2L-1},a_{2L}} \ ,
\end{equation}
where $\chi$ is some tensor of $SU(2)_R \times SU(2)$. These operators need to be renormalized
\begin{equation}
\mathcal{O}_{ren}^M=Z^M_N(\Lambda) \mathcal{O}_{bare}^N \ ,
\end{equation}
where $M$ and $N$ label all the possible operators, $\Lambda$ is an UV cutoff, and $Z$ subtracts all the 
UV divergences from the operator correlator functions. 
The object we are interested in is  the  matrix of anomalous dimensions  $\Gamma$. It is defined as
\begin{equation}\label{gamma}
\Gamma=\frac{d \ln Z}{d \ln \Lambda} \ .
\end{equation}

The eigenstates of $\Gamma$ are conformal operators and the eigenvalues are the corresponding anomalous dimensions.

It is convenient to represent the operators (\ref{op}) as states in a quantum spin chain with $2L$ sites. 
Every site transforms in $(2,2)$ representation of $SU(2)_R \times SU(2)$. The spin chain is alternating 
between the $\mathcal{O}^\dagger$ and $\mathcal{O}$. In this language the mixing matrix (\ref{gamma}) can be 
regarded as the Hamiltonian acting on the Hilbert space $(\bar{V} \otimes V)^{\otimes L}$.  

\section{$SU(2)_R \times SU(2)$ invariant potential}
 
In this section we would like to write  the action (\ref{SN3}) in terms of component fields and in particular we would like to have an explicit expression for the potential in terms of $SU(2)_R \times SU(2)$ invariant objects. We  start by integrating out all the auxiliary fields. In particular the spinorial fields $\chi_{(l)}$ and the bosonic fields $\sigma_{(l)}$, $D_{(l)}$ are all auxiliary fields and  can be eliminated using the equations of motion. From the chiral super fields $Z^I$, $W_j$ we get the complex scalars $Z^i$, $W_j$ and the Dirac spinors $\zeta^i$, $\omega_j$. The potential $V$ can be divided into a part $V^{bos}$ containing only bosonic operators and a part $V^{ferm}$ containing bosonic and fermionic operators. Let consider first the bosonic part.

\subsection{The bosonic potential}
 
The bosonic potential $V^{bos}$ gets  contributions from the superpotential and from the 
Chern-Simons interactions  $V^{bos}= V^{bos}_{W} + V^{bos}_{CS}$. 
The superpotential part is
\begin{equation}
V^{bos}_{W}= \Tr \( \sum_{i,j} \Big|\partial_{Z^i}W \Big|^2+ \Big|\partial_{W_j}W \Big|^2 \) \ ,
\end{equation}
where $W$ is the superpotential given in  eq. (\ref{inter}).
The Chern-Simons part is
\begin{eqnarray}
V^{bos}_{CS}&=&\Tr \left(Z^\dagger_i Z^{i}\sigma^2_{(1)}-2Z^{ i}\sigma_{(1)} Z^\dagger_i\sigma _{(2)}+Z^{ i}Z^\dagger_i\sigma^2_{(2)}\right)\nonumber \\&&+ \Tr\left(W^{\dagger i}W_i\sigma^2_{(2)}-2W_i\sigma_{(2)} W^{\dagger i}\sigma _{(1)}+W_iW^{\dagger i}\sigma^2_{(1)}\right) \ ,
\end{eqnarray}
where
\begin{eqnarray}
\sigma_{(1)}= \frac{2\pi}{k_1} \( Z^\dagger_iZ^{ i}-W_iW^{\dagger i} \) \ ,  \qquad
\sigma_{(2)}= \frac{2\pi}{k_2} \( W^{\dagger i}W_i-W^{ i}W^\dagger_i \) \ .
\end{eqnarray}
If we write a general ansatz by use of  operators in eq. (\ref{OO}) there exist 18 structures compatible with the symmetries and the canonical dimension  of the bosonic fields\footnote{In principle we can write 36 structures which would correspond to the singlets of $SU(2)_R \times SU(2)$. From the group theory computation we get that there are only 25 singlets. It means that there 11 linear relations among the structures. 36 structures are equivalent to 18 different structures modulo cyclic permutation and we find that invariance under cyclic permutation reduces the 11 relations to only 7.}. 

\begin{eqnarray}\label{ansatz}
V^{bos}_{a_n}&& \!\!\!\!\!\!\!\!= a_1\, {\rm Tr} \ O^{a}_iO^{\dagger i}_aO^{b}_jO^{\dagger j}_bO^{c}_kO^{\dagger k}_c+ a_2\, {\rm Tr} \ O^{a}_iO^{\dagger i}_aO^{b}_jO^{\dagger k}_bO^{c}_kO^{\dagger j}_c +a_3\, {\rm Tr} \ O^{a}_iO^{\dagger j}_aO^{b}_kO^{\dagger i}_bO^{c}_jO^{\dagger k}_c  \nonumber\\[0.2cm] 
&& + a_4\, {\rm Tr} \ O^{a}_iO^{\dagger j}_aO^{b}_jO^{\dagger k}_bO^{c}_kO^{\dagger i}_c +a_5\, {\rm Tr} \ O^{a}_iO^{\dagger i}_bO^{b}_jO^{\dagger j}_cO^{c}_kO^{\dagger k}_a+a_6\, {\rm Tr} \ O^{a}_iO^{\dagger i}_bO^{b}_jO^{\dagger k}_cO^{c}_kO^{\dagger j}_a   \nonumber\\ [0.2cm] 
&& +a_7\, {\rm Tr} \ O^{a}_iO^{\dagger j}_bO^{b}_kO^{\dagger i}_cO^{c}_jO^{\dagger k}_a+a_8\, {\rm Tr} \ O^{a}_iO^{\dagger j}_bO^{b}_jO^{\dagger k}_cO^{c}_kO^{\dagger i}_a +a_9\, {\rm Tr} \ O^{a}_iO^{\dagger i}_aO^{b}_jO^{\dagger j}_cO^{c}_kO^{\dagger k}_b \nonumber\\ [0.2cm] 
&&+a_{10}\, {\rm Tr} \ O^{a}_iO{\dagger i}_aO^{b}_jO^{\dagger k}_cO^{c}_kO^{\dagger j}_b +a_{11}\, {\rm Tr} \ O^{a}_iO^{\dagger j}_aO^{b}_kO^{\dagger i}_cO^{c}_jO^{\dagger k}_b+a_{12}\, {\rm Tr} \ O^{a}_iO^{\dagger j}_a  O^{b}_jO^{\dagger k}_cO^{c}_kO^{\dagger i}_b   \nonumber\\ [0.2cm] 
&& +a_{13}\, {\rm Tr} \ O^{a}_iO^{\dagger i}_cO^{b}_jO^{\dagger j}_aO^{c}_kO^{\dagger k}_b+a_{14}\, {\rm Tr} \ O^{a}_iO^{\dagger i}_cO^{b}_jO^{\dagger k}_aO^{c}_kO^{\dagger j}_b +a_{15}\, {\rm Tr} \ O^{a}_iO^{\dagger j}_cO^{b}_kO^{\dagger i}_aO^{c}_jO^{\dagger k}_b   \nonumber\\ [0.2cm] 
&& +a_{16}\, {\rm Tr} \ O^{a}_iO^{\dagger j}_cO^{b}_jO^{\dagger k}_aO^{c}_kO^{\dagger i}_b +a_{17}\, {\rm Tr} \ O^{a}_iO^{\dagger j}_aO^{b}_jO^{\dagger i}_cO^{c}_kO^{\dagger k}_b+a_{18}\, {\rm Tr} \ O^{a}_iO^{\dagger j}_aO^{b}_kO^{\dagger k}_cO^{c}_jO^{\dagger i}_b \nonumber \ , \\
\end{eqnarray}
where $a_n$ are 18 arbitrary real parameters, which we need to fix by use of the explicit expressions  for the bosonic potential in components 
$V^{bos}_{a_n}=V^{bos}$.
If we apply $\dagger$-operation on the ansatz (\ref{ansatz}) we find that the first 16 terms are mapped into themselves, while the last two are mapped into each other. It means that the reality of the potential forces $a_{18}=a_{17}$.
On top of that it appears that some of the 18 structures are linear dependent. If we call $O_n$ the 
operators corresponding to the coefficients $a_n$. We can find the seven linear relations
\begin{eqnarray}\label{depend}
&&3 O_9 -  O_{13} -  O_5 -  O_1=0 \ , \qquad\qquad  \ \ \ 3 O_{12} -  O_{16} -  O_4 -  O_8=0 \ ,\nonumber\\
&&3 O_2 - O_3 -  O_4 -  O_1=0 \ , \qquad\qquad \ \ \  \ 3 O_6 - O_7 - O_5 - O_8=0\ ,\nonumber\\
&&3 O_{14} - O_{16} - O_{15} - O_{13}=0 \ , \qquad\qquad 3 O_{11} - O_3 -  O_7 - O_{15}=0 \ ,\nonumber\\
&&\qquad\qquad\qquad O_{10}- O_9-O_{11}-O_{12}+O_{17}+O_{18}=0 \ .
\end{eqnarray}
In particular, if we try to solve the equation $V^{bos}_{a_n}=V^{bos}$ as a function of the $a_n$ we find a family of solutions parameterized by seven parameters $a_n$ due to the 
relations (\ref{depend}). 
To find the coefficients for the potential we need first to reduce the ansatz by use of (\ref{depend}) to  11 linearly independent structures and then solve  $V^{bos}_{a_n}=V^{bos}$ for the coefficients. This way to proceed means that there are no unique form of the potential if we use the notion of the $SU(2)_R\times SU(2)$-fields. The concrete form of the mixing operator descends from the choice of these 11 structures but the  eigenvalues of the mixing operator are independent of this choice. See additional comments in Appendix \ref{gaugechoice}.   
We found a choice of $a_n$ where 11 of the 18 coefficients are zero.
The remaining non-zero coefficients are
\begin{eqnarray}
&&a_1=-\frac{4\pi^2}{3k_1^2}\ , \qquad a_8=-\frac{4\pi^2}{3k_2^2}\ , \qquad a_{10}=-\frac{8\pi^2}{k_1 k_2}\, \qquad a_{15}=\frac{16\pi^2}{3k_1k_2} \ , \nonumber \\ 
&& \qquad \qquad a_{13}=\frac{16\pi^2(k_1+k_2)}{3k_1^2 k_2}\ , \qquad a_{16}=\frac{16\pi^2(k_1+k_2)}{3k_1k_2^2}  \ .
\end{eqnarray}
In the following we will use these coefficients. The bosonic potential  written in the  explicit $SU(2)_R \times SU(2)$ invariant form is
\begin{eqnarray}\label{potential}
V^{bos} &=& -\frac{4\pi^2}{3k_1^2}\, {\rm Tr} \ O^{a}_iO^{\dagger i}_aO^{b}_jO^{\dagger j}_bO^{c}_kO^{\dagger k}_c-\frac{4\pi^2}{3k_2^2}\, {\rm Tr} \ O^{a}_iO^{\dagger j}_bO^{b}_jO^{\dagger k}_cO^{c}_kO^{\dagger i}_a\nonumber\\
&& -\frac{8\pi^2}{k_1 k_2}\, {\rm Tr} \ O^{a}_iO^{\dagger i}_aO^{b}_jO^{\dagger k}_cO^{c}_kO^{\dagger j}_b +\frac{16\pi^2}{3k_1k_2}\, {\rm Tr} \ O^{a}_iO^{\dagger j}_cO^{b}_kO^{\dagger i}_aO^{c}_jO^{\dagger k}_b\nonumber\\
&&+\frac{16\pi^2(k_1+k_2)}{3k_1^2 k_2}\, {\rm Tr} \ O^{a}_iO^{\dagger i}_cO^{b}_jO^{\dagger j}_aO^{c}_kO^{\dagger k}_b+\frac{16\pi^2(k_1+k_2)}{3k_1k_2^2} \, {\rm Tr} \ O^{a}_iO^{\dagger j}_cO^{b}_jO^{\dagger k}_aO^{c}_kO^{\dagger i}_b\ .\nonumber\\
\end{eqnarray}
With this choice of the coefficients the ABJM limit is apparent. Namely for $k_1+k_2=0$ the last two terms drop out 
and we obtain  the ABJM potential written in $SU(2)_R \times SU(2)$ invariant way. Indeed in this limit the 
R-symmetry and flavor indices of the $O$ fields do not mix anymore due to the R symmetry enhancement to  $SU(4)$ . The remaining coefficients are exactly the ones in \cite{Aharony:2008ug}

\subsection{The fermionic potential}
Le us now proceed with the fermionic potential $V^{ferm}$. Our final goal is to compute the two loops mixing matrix in the planar limit. Part of the contribution to the renormalization of the scalar operators $\mathcal{O}$ in eq. (\ref{op}) comes from fermions running in the loops. This interaction is due to the fermionic potential. The fermionic potential is a quartic function in the fields, each term contains two bosons and two fermions. The contributions are of two types, the first one $V^{ferm}_{ffbb}$ contains terms consisting of two fermions followed by two bosons, the second one $V^{ferm}_{bfbf}$ has the coupling fermions-boson-fermion-boson. It is easy to see that the terms of the second type do not contribute to the mixing matrix at the planar level for the scalar operators. That's why it is enough to consider only the terms of the first type.

The fermionic potential has two contributions, one is coming from the superpotential $V^{ferm}_W$ and the other one coming from the Chern-Simons interactions $V^{ferm}_{CS}$. After integrating out the auxiliary fields we get 
\begin{eqnarray}\label{fermpot}
 V^{{\rm {ferm}}}_W&=&\frac{4\pi }{k_2}\left(\omega_i \zeta^i W_jZ^j+\zeta^i\omega_jZ^jW_i-\zeta^\dagger_i\omega^{\dagger i}Z^\dagger_jW^{\dagger j}-\omega^{\dagger i}\zeta^\dagger_j W^{\dagger j}Z^\dagger_{i}\right)\nonumber\\
&&+\frac{4\pi }{k_1}\left(\omega_i \zeta^j W_jZ^i+\zeta^i\omega_iZ^jW_j-\zeta^\dagger_i\omega^{\dagger j}Z^\dagger_jW^{\dagger i}-\omega^{\dagger i}\zeta^\dagger_i W^{\dagger j}Z^\dagger_{j}\right)+\ldots\nonumber\\
V^{{\rm {ferm}}}_{CS}&=&\frac{2\pi i}{k_1}\left(\zeta^i\zeta^\dagger_i-\omega^{\dagger i}\omega_i\right)\left(Z^jZ_j^\dagger-W^{\dagger j}W_j\right)+ \frac{2\pi i}{k_2}\left(\zeta^\dagger_i\zeta^i-\omega_i \omega^{\dagger i}\right)\left(Z^{\dagger}_jZ^j-W_jW^{\dagger j}\right)\nonumber\\
&&+\frac{4\pi i}{k_1}\left(\zeta^\dagger_i\zeta^j Z_j^\dagger Z^i+\omega_i\omega^{\dagger j}W_j W^{\dagger i}\right) +\frac{4 \pi i}{k_2}\left(\zeta^i\zeta_j^\dagger Z^j Z^\dagger_i+\omega^{\dagger i}\omega_jW^{\dagger j}W_i\right)+\ldots\nonumber\\
\end{eqnarray}
The ellipsis corresponds to couplings in $V^{ferm}_{bfbf}$ which are not relevant for our computation. 
We would like to rewrite the fermionic potential in the $SU(2)_R \times SU(2)$ invariant way. In the ABJM case the superpartners of the scalar field transform in the conjugated representation of the one of the scalars. This is the manifestations of the fact that the $SU(4)$ corresponds to the R-symmetry group of the fields. It means that in the case of the fermionic objects transforming under $SU(2)_R\times SU(2)$ the R-symmetry index should transform in the conjugated representation of the scalar superpartner. However, since we expect that the scalars and spinors belong to the same flavor multiplet they should transform under the same representation of the $SU(2)$ flavor symmetry group. This suggests the following ansatz
%
\begin{eqnarray}\label{fermansatz}
\psi^{\dagger 1i}=-i\zeta^i \ , &\qquad& \psi^{\dagger 2i}=\omega^{\dagger i} \ , \nonumber\\
\psi_{1j}=i\zeta^\dagger_j \ , &\qquad& \psi_{2j}=\omega_j \ .
\end{eqnarray}
The index $i$, $j$ transform under $SU(2)$ flavor symmetry and the written out indices $1,2$ under the $SU(2)_R$-symmetry.
The $SU(2)_R \times SU(2)$ invariant ansatz is then
\begin{eqnarray}
V^{\rm{ferm}}_{f_n}&=&f_1 \Tr\  O^{\dagger i}_a O^a_i \psi^{\dagger bj}\psi_{bj}+f_2  \Tr \ O^{\dagger i}_a O^a_j \psi^{\dagger bj}\psi_{bi}+
f_3  \Tr \ O^{\dagger i}_a O^b_i \psi^{\dagger aj}\psi_{bj}+f_4 \Tr \ O^{\dagger i}_a O^b_j \psi^{\dagger aj}\psi_{bi}\nonumber\\
&&\!\!\!\!\!\!+f_5  \Tr\  O_{i}^{ a} O_a^{\dagger i} \psi_{ bj}\psi^{\dagger bj}+f_6   \Tr \ O_{ i}^{ a} O_a^{\dagger j} \psi_{ bj}\psi^{\dagger bi}+
f_7  \Tr \ O_{ i}^a O^{\dagger i}_b \psi_{ aj}\psi^{\dagger bj}+f_8 \Tr \ O_{ i}^a O^{\dagger j}_b \psi_{ aj}\psi^{\dagger bi} \nonumber\\ &&+\ldots
\end{eqnarray}
The equation $V^{\rm{ferm}}_{f_n}= V^{fer}_W + V^{fer}_{CS}$ gives the solution
\begin{eqnarray}
&&f_1=-\frac{2\pi i}{k_1}\ ,\qquad f_2=0\ ,\qquad f_3=\frac{4\pi i}{k_1}\ , \qquad f_4=\frac{4\pi i}{k_2}\ ,\nonumber\\
&&f_5=-\frac{2\pi i}{k_2}\ ,\qquad f_6=0\ ,\qquad f_7=\frac{4\pi i}{k_2}\ , \qquad f_8=\frac{4\pi i}{k_1} \ .\end{eqnarray}
And the $SU(2)_R \times SU(2)$ invariant fermionic potential is:
\begin{eqnarray}\label{ferpot}
V^{\rm{ferm}}&=&-\frac{2\pi i}{k_1} \Tr O^{\dagger i}_a O^a_i \psi^{\dagger bj}\psi_{bj}+
\frac{4\pi i}{k_1} \Tr O^{\dagger i}_a O^b_i \psi^{\dagger aj}\psi_{bj}+\frac{4\pi i}{k_2} \Tr O^{\dagger i}_a O^b_j \psi^{\dagger aj}\psi_{bi}\nonumber\\
&&-\frac{2\pi i}{k_2} \Tr O_{ i}^a O^{\dagger i}_a \psi_{bj}\psi^{\dagger bj}+
\frac{4\pi i}{k_2} \Tr O_i^a O^{\dagger i}_b \psi_{aj}\psi^{\dagger bj}+\frac{4\pi i}{k_1} \Tr O_i^a O^{\dagger j}_b \psi_{aj}\psi^{\dagger bi}+\ldots\nonumber\\
\end{eqnarray}
The fermionic potential reduces to the ABJM one in the limit $k_1+k_2=0$. Namely, by use of the relation $\delta^i_l\delta_k^j-\delta_k^i\delta^j_l=\epsilon^{ij}\epsilon_{kl}$
and appropriate redefinition of the fields $O^a_i=Y^A,\ \epsilon_{ij}\psi^{\dagger aj}=\psi^{\dagger A}$ where $A$ is an $SU(4)$ index. The last two terms in each line are combined into the terms which mix the $SU(4)$ flavor and the first terms in each line give then flavor non-mixing contributions.
\section{The mixing operator}
\begin{figure}[h]
\begin{center}
  \includegraphics[scale=0.6]{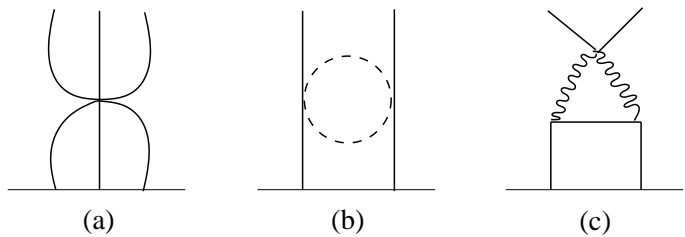}
\caption{\small The graphs that contribute to the mixing operator (a) only scalar bosons are running inside the loops, (b) scalar bosons and fermions are running in the loops, (c) scalar boson and gauge bosons in the loops.}
\label{graphs}
\end{center}
\end{figure}

Right now we have all the tools to compute the dilatation operator $\Gamma$. 
The contributions to the dilatation operator come from the logarithmic divergences ($\ln \Lambda$) of the renormalization function $Z(\Lambda)$. The lowest contributions come at two loops and the non vanishing logarithmic divergences come from the graphs in figure \ref{graphs}. 
The renormalization of the composite operators $\mathcal{O}$ in equation (\ref{op}) comes from three different kind of graphs where (a) only scalar fields, (b) scalar and fermionic fields and  (c) scalar and gauge fields are running in the loops. 
We can analyze them separately. Before doing it let us fix some notation. We are going to compute the Hamiltonian of an $SU(2)_R \times SU(2)$ spin chain in representation $\bf(2,2)$, with alternating sites corresponding to the fields $O$, $O^\dagger$ in the operators $\mathcal{O}$. At every site of the spin chain we have two indices of $SU(2)$ and the final Hamiltonian can be nicely expressed in terms of two basic operators acting on the group indices: the trace operator 
$K: V \otimes \bar{V} \rightarrow  V \otimes \bar{V}$ or $\bar{K}: \bar{V} \otimes V \rightarrow  \bar{V} \otimes V$; and the permutation operator $P: V \otimes V \rightarrow  V \otimes V$ or $P: \bar{V} \otimes \bar{V} \rightarrow  \bar{V} \otimes \bar{V}$. We can distinguish between the operators acting on the R indices ($K$, $P$) and the operators acting on the flavor indices ($\hat{K}, \hat{P}$):
\begin{eqnarray}
&& K^{a'b}_{b'a}= \delta^{a'}_{b'} \delta^{b}_{a}\ ,\qquad  \qquad  K^{i'j}_{j'i}= \delta^{i'}_{j'} \delta^{j}_{i} \ , \nonumber\\
&& P^{a'b'}_{ba}= \delta^{a'}_{b} \delta^{b'}_{a}\ , \qquad  \qquad  P^{i'j'}_{ji}= \delta^{i'}_{j} \delta^{j'}_{i} \ .\nonumber 
\end{eqnarray}
The trace operator $K$ acts on the nearest neighbor sites, while the permutation operator $P$ acts on next to nearest neighbor sites. The 't Hooft couplings $\lambda_i= N/k_i$ are our perturbative expansion parameters. The final expression for the mixing operator $\Gamma$ is a polynomial in $K$ and $P$ with coefficients that are functions of $\lambda_1$, $\lambda_2$. 
\subsection{Six-vertex two-loop diagram}
In this subsection we give the part of the Hamiltonian which comes from the diagram with only scalar fields in the loops.
The graph (a) in figure \ref{graphs} gets the  contribution from the various monomials in the sextic bosonic potential (\ref{potential}). The computation is done in two steps. Firstly, one computes the logarithmic divergent part, and then carefully computes the $SU(2)_R\times SU(2)$ combinatoric structure. 
To write down the final result in a most transparent  way we distinguish between trace operators $\bar K_{l,l+1}$ and $K_{l,l+1}$. 
The first one acts as usual on the sites $\bar V \otimes V$ and gives zero on $V\otimes \bar V$, while the 
second one acts as usual on $V\otimes\bar V$ and gives zero on $\bar V \otimes V$.
The part of the mixing operator coming from this graph is
\begin{eqnarray}
\Gamma_{{\rm {bos}}}&=&\frac{1}{2}\sum_{l=1}^{2L}\Big(-\lambda_1^2{\bar K}_{l,l+1}{\hat{\bar K}}_{l,l+1}-\lambda_2^2{ K}_{l,l+1}{\hat{K}}_{l,l+1}+2\lambda_1\lambda_2 P_{l,l+2}\hat P_{l,l+2}\nonumber\\&&
-\lambda_1\lambda_2\big(1\hat 1+K_{l,l+1}P_{l,l+2} \ \hat{K}_{l,l+1}\hat{P}_{l,l+2} +{\bar K}_{l,l+1}P_{l,l+2} \ \hat{\bar K}_{l,l+1}\hat{P}_{l,l+2} \nonumber\\&&\qquad\qquad \ \  +P_{l,l+2}K_{l,l+1} \ \hat{P}_{l,l+2}\hat{K}_{l,l+1} +P_{l,l+2}{\bar K}_{l,l+1} \ \hat{P}_{l,l+2}\hat{\bar K}_{l,l+1}\big)\nonumber\\&&+4(\lambda_1\lambda_2+\lambda_1^2)P_{l,l+2}\hat{\bar K}_{l,l+1}+4(\lambda_1\lambda_2+\lambda_1^2)P_{l,l+2}\hat{ K}_{l,l+1}\Big) \ .
\end{eqnarray}
\subsection{Fermionic contribution}
The fermionic potential (\ref{ferpot}) gives two types of contributions to the graph (b) in figure \ref{graphs}:
a contribution proportional to the identity in the $SU(2)_R\times SU(2)$ indices, namely a vacuum energy contribution coming from the first two monomials in the two lines of (\ref{ferpot}) and an interacting contribution 
containing the $K$, $\hat{K}$ trace operators. The constant part of the full mixing matrix gets contribution also 
from other graphs than the ones in figure \ref{graphs}, for example, from the renormalization to the propagator $<O^\dagger O>$. We are not going to compute these diagrams. Later, we fix this constant part using supersymmetry. For this reason we concentrate here only on the contributions coming from the last two monomials in each lines in (\ref{ferpot}). After computing the logarithmic divergent part of the graph (b) in figure \ref{graphs} and computing the combinatorial $SU(2)_R\times SU(2)$ structure, we obtain the fermionic contribution to the mixing operator
\begin{eqnarray}
\Gamma_{\rm {ferm}}&=&\sum_{l=1}^{2L}\Big(2(\lambda_2^2+\lambda_1\lambda_2)\bar K_{l,l+1}\hat 1+\lambda_1^2 \bar K_{l,l+1}\hat{\bar K}_{l,l+1} +2(\lambda_1^2+\lambda_1\lambda_2) K_{l,l+1}\hat 1+\lambda_2^2  K_{l,l+1}\hat{K}_{l,l+1}\Big) \ .\nonumber\\
\end{eqnarray}
\subsection{The gauge bosons contribution}
The last contribution to the mixing operator comes from the graph (c) in figure \ref{graphs}. 
The gauge bosons do not carry $SU(2)_R\times SU(2)$ indices and we just need to compute the 
two loop diagram with the correct coupling constants coming from the scalar-gauge interactions in the Lagrangian.   
The final result is
\begin{equation}
\Gamma_{\rm {gauge}}=-\frac{1}{2}\sum_{l=1}^{2L}\Big(\lambda_2^2{\bar  K}_{l,l+1} \hat {\bar K}_{l,l+1}+\lambda^2_1{ K}_{l,l+1} \hat { K}_{l,l+1}\Big) \ .
\end{equation}
\subsection{Two-loop dilatation operator}
The complete two loop mixing operator is obtained summing up $\Gamma_{{\rm {bos}}}$, $\Gamma_{\rm {ferm}}$ and $\Gamma_{\rm {gauge}}$. Before writing down the final expression we need to fix the constant contribution. Supersymmetry implies that the anomalous dimension of the symmetric traceless operators is equal to zero. This fact fixes the constant contribution. The complete\footnote{We would like to stress here that since there are relations between the trace and permutation operators acting on two-dimensional indices the above form of the Hamiltonian is not unique. The action of the Hamiltonian is of course independent of the concrete representation in terms of $K$s and $P$s.} Hamiltonian can be written as
\begin{eqnarray}\label{hful}
\Gamma_{{\rm {full}}}&=&\frac{1}{2}\sum_{l=1}^{2L}\Big((\lambda_1^2-\lambda_2^2){\bar K}_{l,l+1}{\hat{\bar K}}_{l,l+1}+(\lambda_2^2-\lambda_1^2){ K}_{l,l+1}{\hat{K}}_{l,l+1}\nonumber\\
&&+4(\lambda_1\lambda_2+\lambda_1^2)(P_{l,l+2}\hat{\bar K}_{l,l+1}+ K_{l,l+1}\hat 1)+4(\lambda_1\lambda_2+\lambda_2^2)(P_{l,l+2}\hat{ K}_{l,l+1}+\bar K_{l,l+1} \hat 1)\nonumber
\\&&
-\lambda_1\lambda_2\big(2-2 P_{l,l+2}\hat P_{l,l+2}+K_{l,l+1}P_{l,l+2} \ \hat{K}_{l,l+1}\hat{P}_{l,l+2} +{\bar K}_{l,l+1}P_{l,l+2} \ \hat{\bar K}_{l,l+1}\hat{P}_{l,l+2} \nonumber\\&&\qquad\qquad \ \  +P_{l,l+2}K_{l,l+1} \ \hat{P}_{l,l+2}\hat{K}_{l,l+1} +P_{l,l+2}{\bar K}_{l,l+1} \ \hat{P}_{l,l+2}\hat{\bar K}_{l,l+1}\big)\Big) \ .\nonumber\\
\end{eqnarray}
The last two lines are the only contributions to the mixing operator in the ABJM case.
Indeed in the limit $k_1+k_2=0$ the Hamiltonian reduces to
\begin{eqnarray}
\Gamma_{{\rm {full}}}^{ABJM}&=&\frac{\lambda^2}{2}\sum_{l=1}^{2L}\Big(2-2 P_{l,l+2}\hat P_{l,l+2}+K_{l,l+1}P_{l,l+2} \ \hat{K}_{l,l+1}\hat{P}_{l,l+2} + P_{l,l+2}K_{l,l+1} \ \hat{P}_{l,l+2}\hat{K}_{l,l+1}\Big) \ \nonumber\\
\end{eqnarray}
that is exactly the mixing operator in \cite{Minahan:2008hf} written in $SU(2)_R\times SU(2)$ invariant form, where we didn't distinguish between $K$, $P$ and $\bar{K}$, $\bar{P}$. 

It is nice to observe that one can define a parity operator $\mathcal{P}$ acting on the spin chain. Its action reverses the orientation of the chain from clockwise to anticlockwise or vice versa. In particular it acts on the operators as $$\mathcal{P}\hbox{  } \Tr \( O^{\dagger i_1}_{a_1} O^{a_2}_{i_2}...O^{\dagger i_{2L-1}}_{a_{2L-1}} O^{a_{2L}}_{i_{2L}}\) = \Tr \( O^{a_{2L}}_{i_{2L}} O^{\dagger i_{2L-1}}_{a_{2L-1}}...O^{a_{2}}_{i_{2}} O^{\dagger i_{1}}_{a_{1}}\)\ .$$ The parity operation\footnote{If we act with the  parity operator on the Hamiltonian the transformed one should act on the parity transformed states as the original Hamiltonian on the non transformed states. The new vertices of a such transformed Hamiltonian are obtained from the full potential by acting on all the terms with the parity operator. This corresponds exactly to the exchange of $\lambda_1$ and $\lambda_2$ in eq. (\ref{hful}) or alternatively to the exchange of $K,\hat K$ and $\bar K,\hat {\bar K}$.} on the Hamiltonian (\ref{hful}) exchanges $\lambda_1$ and $\lambda_2$. The parity transformed Hamiltonian is
\begin{eqnarray}
\mathcal{P} \hbox{  }\Gamma_{{\rm {full}}}\ \mathcal{P}&=&\frac{1}{2}\sum_{l=1}^{2L}\Big((\lambda_2^2-\lambda_1^2){\bar K}_{l,l+1}{\hat{\bar K}}_{l,l+1}+(\lambda_1^2-\lambda_2^2){ K}_{l,l+1}{\hat{K}}_{l,l+1}\nonumber\\
&&+4(\lambda_1\lambda_2+\lambda_2^2)(P_{l,l+2}\hat{\bar K}_{l,l+1}+ K_{l,l+1}\hat 1)+4(\lambda_1\lambda_2+\lambda_1^2)(P_{l,l+2}\hat{ K}_{l,l+1}+\bar K_{l,l+1} \hat 1)\nonumber
\\&&
-\lambda_1\lambda_2\big(2-2 P_{l,l+2}\hat P_{l,l+2}+K_{l,l+1}P_{l,l+2} \ \hat{K}_{l,l+1}\hat{P}_{l,l+2} +{\bar K}_{l,l+1}P_{l,l+2} \ \hat{\bar K}_{l,l+1}\hat{P}_{l,l+2} \nonumber\\&&\qquad\qquad \ \  +P_{l,l+2}K_{l,l+1} \ \hat{P}_{l,l+2}\hat{K}_{l,l+1} +P_{l,l+2}{\bar K}_{l,l+1} \ \hat{P}_{l,l+2}\hat{\bar K}_{l,l+1}\big)\Big)\nonumber \ .\\
\end{eqnarray}
For $\lambda_1\neq \pm \lambda_2$ the parity symmetry of the Hamiltonian is broken by the terms in the first and second line. The only values of $\lambda_1$ and $\lambda_2$ which correspond to the parity invariant Hamiltonian are $\lambda_1=\pm\lambda_2$. 

\section{Length four operators}

A typical sign of integrability of a system is the presence of different operators with the same anomalous dimensions. \cite{Beisert:2003tq,Kristjansen:2008ib} In the ABJM case this happens for example for operators of length four \cite{Minahan:2008hf}. In that case the system is an $SU(4)$ spin chain alternating between fundamental $\bf 4$ representation and antifundamental $\bf \bar{4}$ representation. The $\bf 4$ is associated with the vector: $Y^A=(Z^1,Z^2,W^\dagger_1,W^\dagger_2)$ and the length four operators are
 $\Tr \left( Y^{A_1} Y^\dagger_{B_1}Y^{A_2} Y^\dagger_{B_2} \right) $. If we decompose these operators in representations of $SU(4)$ we find that they contain two singlets $\bf 1$, two adjoints $\bf 15$, one $\bf 20$ and one $\bf 84$ representations. It happens that the two adjoint operators have the same anomalous dimension $6\lambda^2$. The natural question is, what happens to these operators in the case in which $k_1 \ne - k_2$? Are they still degenerate? To answer these questions we consider the following operators
\begin{equation}\label{ofour}
{\rm Tr}\ O^{\dagger i_1}_{a_1} O^{a_2}_{i_2} O^{\dagger i_3}_{a_3} O^{a_4}_{i_4} \ .
\end{equation}
They decompose in representation of $SU(2)_R \times SU(2)$. In particular the $\bf 15$ of $SU(4)$ 
decomposes under $SU(2)_R\times  SU(2)$ as
\begin{equation} 
\bf 15 \rightarrow  (3,1) + (1,3) + (3,3) \ . \nonumber
\end{equation}
For this reason in this section we will be interested to apply the Hamiltonian (\ref{hful}) to operators in (\ref{ofour}) in representations $\bf (3,1)$, $\bf (1,3)$ and $\bf (3,3)$. Operators with the same quantum numbers typically mix among each other under renormalization. We need to consider all the operators of the same length that transform in the same representation. The operators in the representation $\bf (3,1)$ and $\bf (1,3)$ come only from the decomposition of the $\bf 15$ of $SU(4)$, but there exist other three operators in the $\bf (3,3)$ representation coming respectively: one from the $\bf 20$ and two from the $\bf 84$. As result we have two operators in the $\bf (3,1)$, two in the $\bf (1,3)$, and five in the $\bf (3,3)$. In the following subsections we are going to analyze separately their anomalous dimensions and to check if the degeneracy which is present in the integrable ABJM case is still there or is lifted.

\subsection{Operators in (3,1)} 

Let us start with the operators in representation $({\bf 3,1})$. From the decomposition in the list (\ref{decomposition}) in the Appendix we know that there are six structures transforming in the representation ({\bf 3,1}), four come from {\bf 15} and two from {\bf 45} and ${\bf \overline {45}}$ of $SU(4)$. Only the structures descending from the {\bf 15} of $SU(4)$ can form operators invariant under trace. Indeed cyclicity relates four states and we get just two operators:
\begin{eqnarray}\label{op31}
{\rm Tr}\ |1-{ \bf 15}\rangle_{\bf (3,1)}&=&{\rm Tr}\ O^{\dagger i}_a O^a_i O^{\dagger m}_b O^c_m-{\rm trace} \ , \nonumber\\
{\rm Tr}\   |2-{\bf 15}\rangle_{\bf (3,1)}&=&{\rm Tr}\ O^{\dagger m}_b O^a_i O^{\dagger i}_a O^c_m-{\rm trace} \ .
\end{eqnarray}
The first label enumerates the operators and the second one gives the corresponding $SU(4)$ multiplet.

Applying the mixing operator we obtain
\small
\begin{eqnarray}\label{31one}
\Gamma \ {\rm Tr}\ |1-{\bf 15}\rangle_{\bf (3,1)}&=&2(\lambda_1^2-\lambda_1\lambda_2+\lambda_2^2){\rm Tr}\ |1-{\bf 15}\rangle_{\bf (3,1)}+(5\lambda_2-\lambda_1)(\lambda_1+\lambda_2){\rm Tr}\ |2-{\bf 15}\rangle_{\bf (3,1)}\nonumber\\
&&+6\lambda_2(\lambda_1+\lambda_2)\Big({\rm Tr}\ O^{\dagger i}_a O^c_i O^{\dagger j}_b O^a_j+{\rm Tr}\ O^{\dagger i}_b O^a_i O^{\dagger j}_a O^c_j\Big)\nonumber\\[0.25cm]
&=&2(\lambda_2^2+5\lambda_1\lambda_2+7\lambda_1^2){\rm Tr}\ |1-{\bf 15}\rangle_{\bf (3,1)}+(5\lambda_2-\lambda_1)(\lambda_1+\lambda_2){\rm Tr}\ |2-{\bf 15}\rangle_{\bf (3,1)}\nonumber
\end{eqnarray}
\begin{eqnarray}\label{31two}
\Gamma\ {\rm Tr}\ |2-{\bf 15}\rangle_{\bf (3,1)}&=&2(\lambda_1^2-\lambda_1\lambda_2+\lambda_2^2){\rm Tr}\ |2-{\bf 15}\rangle_{\bf (3,1)}+(5\lambda_1-\lambda_2)(\lambda_1+\lambda_2){\rm Tr} \ |1-{\bf 15}\rangle_{\bf (3,1)}\nonumber\\
&&+6\lambda_1(\lambda_1+\lambda_2)\Big({\rm Tr}\ O^{\dagger i}_a O^a_j O^{\dagger j}_b O^c_i+{\rm Tr}\ O^{\dagger i}_b O^c_j O^{\dagger j}_a O^a_i\Big)\nonumber\\[0.25cm]
&=&2(\lambda_1^2+5\lambda_1\lambda_2+7\lambda_2^2){\rm Tr}\ |2-{\bf 15}\rangle_{\bf (3,1)}+(5\lambda_1-\lambda_2)(\lambda_1+\lambda_2){\rm Tr}\ |1-{\bf 15}\rangle_{\bf (3,1)}\nonumber
\end{eqnarray}
\normalsize

The application of the mixing  operator on the states ${\rm Tr}\ |1-\bf 15\rangle_{\bf (3,1)}$ and  ${\rm Tr}\ |2- \bf 15\rangle_{\bf (3,1)}$ produces structures which we cannot immediately match with the basis states. This comes from the fact that there are more structures than the linearly independent ones. There are 6 ways to organize the R-symmetry indices in such a way that they transform in representation {\bf 3} of $SU(2)_R$ and two ways to organize the flavor indices that transform in {\bf 1} of $SU(2)$. Using the relations from Appendix B these 12 structures can be related to the 6 basis structures which come from the decomposition of {\bf 15 , 45 } and { $\overline {\bf 45}$} of $SU(4)$. The eigenvalues are
\begin{equation}
8\lambda_1^2+10\lambda_1\lambda_2+8\lambda_2^2\pm (\lambda_1+\lambda_2)\sqrt{31 \lambda_1^2-46 \lambda_1\lambda_2 +31\lambda_2^2} \ .
\end{equation}
For physical real values of $\lambda_1$, $\lambda_2$ the eigenvalues are degenerate only for $\lambda_1=-\lambda_2 = \lambda$. In this case our result reduces to the ABJM one \cite{Minahan:2008hf} and the two operators in (\ref{op31}) have the same  anomalous dimension, $6\lambda^2$. In all the other cases the degeneracy is lifted.


\subsection{Operators in (1,3)}

The operators in representation $\bf(1,3)$, similarly to the previous case, appear also in the decomposition of the {\bf 15} of $SU(4)$. As in the $\bf(3,1)$, we get only two operators
\begin{eqnarray}
{\rm Tr} \ |1-\bf 15\rangle_{\bf (1,3)}&=& {\rm Tr} \ O^{\dagger i}_a O^a_i O^{\dagger j}_b O^b_k-{\rm trace} \ , \nonumber\\
{\rm Tr} \ |2- \bf 15\rangle_{\bf (1,3)}&=&{\rm Tr} \ O^{\dagger j}_b O^a_i O^{\dagger i}_a O^b_k-{\rm trace} \ .
\end{eqnarray}
Again using the relations from the Appendix B we obtain
\begin{eqnarray}
 \Gamma \ {\rm Tr} \ |1-{\bf 15}\rangle_{\bf (1,3)}&=&2(3\lambda_1^2-\lambda_1\lambda_2+\lambda_2^2){\rm Tr}\ |1-{\bf 15}\rangle_{\bf (1,3)}\nonumber\\
 &&+(\lambda_1+\lambda_2)(5\lambda_1+7\lambda_2){\rm Tr}\ |2- {\bf 15}\rangle_{\bf (1,3)}\ , \nonumber\\[0.2cm]
 \Gamma \ {\rm Tr} \ |2-{\bf 15}\rangle_{\bf (1,3)}&=&2(3\lambda_2^2-\lambda_1\lambda_2+\lambda_1^2){\rm Tr}\ |2-{\bf 15}\rangle_{\bf (1,3)}\nonumber\\
 &&+(\lambda_1+\lambda_2)(5\lambda_2+7\lambda_1){\rm Tr}\ |1- {\bf 15}\rangle_{\bf (1,3)}\ . 
\end{eqnarray}
The eigenvalues are:
\begin{equation}
2(2\lambda_1^2+\lambda_1\lambda_2+2\lambda_2^2) \pm (\lambda_1+\lambda_2)\sqrt{3(13\lambda_1^2+22\lambda_1\lambda_2+13 \lambda^2_2)}.
\end{equation}
As in the previous case the mixing and the anomalous dimensions reduce to the ABJM ones \cite{Minahan:2008hf} in the limit $\lambda_1=-\lambda_2$, otherwise the degeneracy is lifted.

\subsection{Operators in (3,3)}

The $\bf (3,3)$ case is a bit more involved. As we can see in  the list (\ref{decomposition}) there are nine structures transforming in $\bf (3,3)$ which come from the decomposition of the length four structures of $SU(4)$. 
Two of them coming from {\bf 45} and $\overline {\bf 45}$, due to the antisymmetrization, do not correspond to any operators. From the remaining seven structures the four coming from {\bf 15} of $SU(4)$ correspond to two trace invariant operators. Altogether we have the following basis for the operators in $\bf (3,3)$.\footnote{In principle we can write two operators which would correspond to the decomposition of {\bf 20}, the one with upper indices symmetrized and lower antisymmetrized and vice versa. 
By use of the relations in Appendix \ref{rel33} one can show that one of these two structures can be written as a linear combination of the remaining one, ${\rm Tr}\ |1-{\bf 15}\rangle_{\bf (3,3)}$ and ${\rm Tr}\ |2-{\bf 15}\rangle_{\bf (3,3)}$.} 
\begin{eqnarray}\label{basisoperators}
{\rm Tr}\ |1-{\bf 15}\rangle_{\bf (3,3)}&=& {\rm Tr}\ O^{\dagger i}_a O^a_i O^{\dagger j}_b O^c_k-{\rm trace} \ ,\nonumber\\[0.15cm]
{\rm Tr}\ |2-{\bf 15}\rangle_{\bf (3,3)}&=&{\rm Tr}\ O^{\dagger j}_b O^a_i O^{\dagger i}_a O^c_k-{\rm trace} \ ,\nonumber\\[0.15cm]
{\rm Tr}\ |3-{\bf 20}\rangle_{\bf (3,3)}&=&{\rm Tr}\Big( O^{\dagger [i}_{(b} O^{[a}_{(k} O^{\dagger l]}_{e)} O^{d]}_{m)}-{\rm traces}\Big)\epsilon_{ad}\epsilon^{ce}\epsilon_{il}\epsilon^{jm}\nonumber\\&=&4\ {\rm Tr } \ O^{\dagger [i}_{(b} O^{[a}_{(k} O^{\dagger j]}_{a)} O^{c]}_{i)} -{\rm Tr}\ |1-{\bf 15}\rangle_{\bf (3,3)}+{\rm Tr}\ |2-{\bf 15}\rangle_{\bf (3,3)}- {\rm traces}\ ,\nonumber\\[0.15cm]
{\rm Tr}\ |4-{\bf 84}\rangle_{\bf (3,3)}&=&{\rm Tr}\ \Big(O^{\dagger (j}_{(b} O^{[a}_{[i} O^{\dagger m)}_{e)} O^{d]}_{l]}-{\rm traces}\Big)\epsilon_{ad}\epsilon^{ce}\epsilon^{il}\epsilon_{km}\nonumber\\
&=&4\ {\rm Tr } \ O^{\dagger (j}_{(b} O^{[a}_{[i} O^{\dagger i)}_{a)} O^{c]}_{k]} -\frac{1}{3} \ {\rm Tr}\ |1-{\bf 15}\rangle_{\bf (3,3)}-\frac{1}{3}{\rm Tr}\ |2-{\bf 15}\rangle_{\bf (3,3)}- {\rm traces}\ , \nonumber\\[0.15cm]
{\rm Tr}\ |5-{\bf 84}\rangle_{\bf (3,3)}&=&{\rm Tr}\ \Big(O^{\dagger [i}_{[a} O^{(c}_{(k} O^{\dagger l]}_{d]} O^{e)}_{m)}-{\rm traces}\Big)\epsilon^{ad}\epsilon_{be}\epsilon_{il}\epsilon^{jm} \nonumber\\
&=&4\ {\rm Tr } \ O^{\dagger [j}_{[b} O^{(a}_{(i} O^{\dagger i]}_{a]} O^{c)}_{k)} -\frac{1}{3} \ {\rm Tr}\ |1-{\bf 15}\rangle_{\bf (3,3)}-\frac{1}{3}{\rm Tr}\ |2-{\bf 15}\rangle_{\bf (3,3)}- {\rm traces}\ . \nonumber\\
\end{eqnarray}
The first number enumerates the operators and the second one gives the representation of $SU(4)$ to which it corresponds. 
The states  ${\rm Tr}\ |1-{\bf 15}\rangle_{\bf (3,3)}$ and ${\rm Tr}\ |2-{\bf 15}\rangle_{\bf (3,3)}$  in the definition of the last three operators come from the decomposition of the traces of the $SU(4)$ operators, ${\bf 20}$ and ${\bf 84}$.
To obtain the mixing matrix of anomalous dimensions we apply the Hamiltonian (\ref{hful}) to the above basis states. In general the result will contain  structures which do not match with the five basis operators in the list(\ref{basisoperators}). We used  the relations listed in Appendix \ref{rel33}.
The mixing matrix is

\footnotesize
\begin{equation}
\left(
\begin{array}{ccccc}
\frac{2}{3}\left(7\lambda_1^2+3\lambda_1\lambda_2+5 \lambda_2^2\right) & \frac{1}{3}\left(\lambda_1+\lambda_2\right)\left(7\lambda_1+5\lambda_2\right) &0 &-\frac{8}{3}\lambda_1(\lambda_1+\lambda_2) &-\frac{8}{3}\lambda_1(\lambda_1+\lambda_2)\\
 \frac{1}{3}\left(\lambda_1+\lambda_2\right)\left(5\lambda_1+7\lambda_2\right) &\frac{2}{3}\left(5\lambda_1^2+3\lambda_1\lambda_2+7 \lambda_2^2\right) &0 &-\frac{8}{3}\lambda_2(\lambda_1+\lambda_2) &-\frac{8}{3}\lambda_2(\lambda_1+\lambda_2)\\
0&0&2(\lambda_1-\lambda_2)^2&2(\lambda_1^2-\lambda_2^2)&-2(\lambda_1^2-\lambda_2^2)\\
-(\lambda_1+\lambda_2)(2\lambda_1+\lambda_2)&-(\lambda_1+\lambda_2)(\lambda_1+2\lambda_2)&-\lambda_1^2+\lambda_2^2&3(\lambda_1+\lambda_2)^2&(\lambda_1+\lambda_2)^2\\
-(\lambda_1+\lambda_2)(2\lambda_1+\lambda_2)&-(\lambda_1+\lambda_2)(\lambda_1+2\lambda_2)&-\lambda_1^2+\lambda_2^2&(\lambda_1+\lambda_2)^2&3(\lambda_1+\lambda_2)^2
\end{array}\right)\nonumber
\end{equation}
\normalsize

In the ABJM-limit, $\lambda_1=-\lambda_2=\lambda$, the eigenstates and their corresponding eigenvalues are \cite{Minahan:2008hf} 
\begin{eqnarray}
  {\rm Tr} \ |1-\bf 15\rangle_{\bf (3,3)} \ :&& 6\lambda^2 \ , \nonumber\\
 {\rm Tr} \ |2-\bf 15\rangle_{\bf (3,3)} \ : && 6\lambda^2  \ , \nonumber\\
{\rm Tr} \ |3-\bf 20\rangle_{\bf (3,3)}  \ :&& 8\lambda^2 \ , \nonumber \\
 {\rm Tr} \ |4-\bf 84\rangle_{\bf (3,3)}  \ :&& 0 \ ,   \nonumber\\
 {\rm Tr} \ |5-\bf 84\rangle_{\bf (3,3)}  \ :&& 0  \ .
\end{eqnarray}
There are other particular values of $\lambda_1$, $\lambda_2$. For $\lambda_1=\lambda_2$ the theory is still parity invariant, but we don't find any degeneracy pairs among the eigenstates which would map one into each other  under the parity transformation. For the values of $\lambda_1$, $\lambda_2$ outside the regime $\lambda_1\neq-\lambda_2$ we can find  degeneracy among the eigenvalues of the mixing matrix, but since the theory is not parity invariant the operators with the same anomalous dimensions do not form parity pairs. These results suggest that the ABJM integrability is broken for generic values of $\lambda_1$ and $\lambda_2$.
\subsection{Integrability and degeneracy}
Let us try to get some conclusions related to the integrability of the system. As we claimed at the beginning of this section a generic feature of integrability is the presence of degeneracy pairs \cite{Beisert:2003tq,Kristjansen:2008ib}. Namely, the existence of couples of operators which have the same anomalous dimension and which are mapped one into each other by the parity operator $\mathcal{P}$. In the ABJM spin chain the first example of degeneracy pairs is in the set of length four operators: they are the operators in the adjoint representation of $SU(4)$. In this section we checked that all the $SU(2)_R\times  SU(2)$ operators which are contained in the decomposition of the ABJM degeneracy pairs are no longer degeneracy pairs for generic $k_1$, $k_2$. This fact could be interpreted as a weak evidence of the absence of integrability of the system for $k_1 \ne -k_2$. 

Let us explain why this is just a weak evidence. First of all the parity symmetry is broken by the Hamiltonian (\ref{hful}) for generic values of $k_1$, $k_2$. A nice observation is that parity is restored for $k_1=\pm k_2$. One of these two points is the ABJM limit where degeneracy pairs appear and the system is integrable. The other point is still parity invariant but there is no degeneracy in the anomalous dimensions. Even this observation is not conclusive: the original eigenvectors of the ABJM mixing matrix are no longer eigenvectors of the new Hamiltonian. The new eigenvectors do not form pairs under parity,  they are actually parity eigenvectors and we cannot claim that integrability is broken because they do not have the same anomalous dimension. To say something stronger about the integrability of the theory one should compute for example the mixing of longer operators, or directly compute the integrable Hamiltonian associated to the $SU(2)_R\times  SU(2)$ spin chain, but
 also in this case the claim could be not definitive.

Even with all these subtleties in mind we would like to take the lifting of the degeneracy, which is present in the ABJM limit, as a hint against the integrability of the system. Of course, a more rigorous analysis is required.
\section{Conclusions}
In this note we started the analysis of the deformed integrable Chern-Simons theories. As a first example we considered  the ABJM theory with arbitrary Chern-Simons levels $k_1$, $k_2$. We constructed the complete two loop mixing operator for the bosonic scalar sector of the theory and we computed the anomalous dimension for some length four operators. We observed that the degeneracy of anomalous dimensions  which is present in the integrable limit (the ABJM theory) disappears for generic $k_1$and $k_2$. We interpreted this fact as a weak evidence of the absence of integrability for these theories, namely, when $k_1 + k_2 \ne 0$ the ABJM integrability seems to be destroyed.
A possible future direction could be to start a deeper investigation of the integrability of these theories, in field theory and maybe in the IIA string dual, to support or contradict our conclusions.

Another  nice application of the ideas presented in this note could be a more general analysis of the integrability of Chern-Simons quiver gauge theories. For example it would be nice to see what happens to the integrable properties of Chern-Simons theories that come by orbifolding ABJM, once we allow non orbifold values for the various $k_i$. We hope to come back to this problem in the near future.

We hope to have convinced the reader that three dimensional Chern-Simons theories are a nice laboratory to study integrability, and in a sense, due to the quartic interactions and the presence of Chern-Simons levels, they allow a perturbative weak coupling analysis of more general deformations than the four dimensional examples.

\section*{Acknowledgments}
We are happy to thank first of all Konstantin Zarembo for many nice discussions, and Claudio Destri, Giuseppe Policastro, Alessandro Tomasiello, Jan Troost, Alberto Zaffaroni, Andrey Zayakin for valuable conversations. W.S. would like to thank Ludwig-Maximilians-University and University of Hamburg for kind hospitality where part of this work was done.
D. ~F.~ is supported by CNRS and ENS Paris. The research of W.~S. was supported in part by the ANR (CNRS-USAR) contract 05-BLAN-0079-01.

\appendix
\appendixpage
\section{Relations among the operator structures of (3,3)}\label{rel33}
If we hold one type of the coefficients fixed we can can write down six structures corresponding to representation {\bf 3} of the other type of coefficients.

\begin{eqnarray}\label{sixstructures}
|1\rangle_{\bf 3} =&&O^\dagger_aO^aO^\dagger_b O^c - {\rm trace}\nonumber\\
|2\rangle_{\bf 3} =&&O^\dagger_bO^aO^\dagger_a O^c - {\rm trace}\nonumber\\
|3\rangle_{\bf 3} =&&O^\dagger_bO^cO^\dagger_a O^a - {\rm trace}\nonumber\\
|4\rangle_{\bf 3} =&&O^\dagger_aO^cO^\dagger_b O^a - {\rm trace}\nonumber\\
|5\rangle_{\bf 3} =&&O^\dagger_{(b}O^{a}O^\dagger_{e)} O^{d}\epsilon_{ad}\epsilon^{ce}\nonumber\\
|6\rangle_{\bf 3} =&&O^\dagger_{a}O^{(c}O^\dagger_{d} O^{e)}\epsilon^{ad}\epsilon_{be}
\end{eqnarray}

From the group theory computation we know that there should be only three  independent structures transforming in the representation {\bf 3}.
\begin{equation}
{\bf 2}\otimes {\bf 2}\otimes {\bf 2}\otimes {\bf 2}= {\bf 1}^2\oplus {\bf 3}^3 \oplus {\bf {5}}
\end{equation}
If we consider the following relations
\begin{eqnarray}
O^\dagger_bO^cO^\dagger_aO^a&=&\epsilon_{be}\epsilon^{ad}O^{\dagger e}O^c O^\dagger_aO_d\nonumber\\
&=&\left(\delta^d_b\delta_e^a-\delta^a_b\delta^d_e\right)O^{\dagger e}O^cO^\dagger_aO_d\nonumber\\
&=&O^{\dagger a}O^cO^\dagger_a O_b-O^{\dagger d}O^cO^\dagger_bO_d\nonumber\\
&=&\epsilon^{ad}\epsilon_{be}O^\dagger_d O^c O^\dagger_a O^e+O^\dagger_a O^c O^\dagger_b O^a\nonumber
\end{eqnarray}
\begin{eqnarray}
O^\dagger_bO^cO^\dagger_aO^a&=&\epsilon_{ae}\epsilon^{cd}O^{\dagger }_bO_d O^{\dagger e}O^a\nonumber\\
&=&\left(\delta^c_e\delta_a^d-\delta^c_a\delta^d_e\right)O^{\dagger }_bO_dO^{\dagger e}O^a\nonumber\\
&=&O^{\dagger }_bO_aO^{\dagger c} O^a-O^{\dagger }_bO_aO^{\dagger a}O^c\nonumber\\
&=&\epsilon_{ad}\epsilon^{ce}O^\dagger_b O^d O^\dagger_e O^a+O^\dagger_b O^a O^\dagger_a O^c\nonumber
\end{eqnarray}
\begin{eqnarray}
O^\dagger_aO^bO^\dagger_c O^a &=&\epsilon_{ad}\epsilon^{be}O^{\dagger d}O_e O^\dagger_c O^a\nonumber\\
&=&\left(\delta^e_a\delta^b_d-\delta_d^e\delta_a^b\right)O^{\dagger d}O_e O^{\dagger}_c O^a\nonumber\\
&=&O^{\dagger b}O_a O^\dagger_c O^a-O^{\dagger a}O_a O^\dagger_c O^b\nonumber\\
&=&\epsilon^{bd}\epsilon_{ae}O^\dagger_d O^eO^\dagger_c O^a+O^{\dagger }_aO^a O^\dagger_c O^b\nonumber
\end{eqnarray}
\begin{eqnarray}\label{rel1}
O^\dagger_bO^aO^\dagger_a O^c &=&\epsilon^{ad}\epsilon_{be}O^{\dagger e}O_d O^\dagger_a O^c\nonumber\\
&=&\left(\delta^a_e\delta^d_b-\delta_b^a\delta_e^d\right)O^{\dagger e}O_d O^{\dagger}_a O^c\nonumber\\
&=&O^{\dagger a}O_b O^\dagger_a O^c-O^{\dagger a}O_a O^\dagger_b O^c\nonumber\\
&=&\epsilon^{ad}\epsilon_{be}O^\dagger_d O^eO^\dagger_a O^c+O^{\dagger }_aO^a O^\dagger_b O^c
\end{eqnarray}
we can find the following set of the relations among the six structures listed in (\ref{sixstructures})
\begin{eqnarray}\label{rel}
&&|1\rangle_{\bf 3}+|2\rangle_{\bf 3}-|3\rangle_{\bf 3}-|4\rangle_{\bf 3}-2|5\rangle_{\bf 3}=0\nonumber\\
&&|1\rangle_{\bf 3}-|2\rangle_{\bf 3}-|3\rangle_{\bf 3}+|4\rangle_{\bf 3}-2|6\rangle_{\bf 3}=0\nonumber\\
&&|1\rangle_{\bf 3}-|2\rangle_{\bf 3}+|3\rangle_{\bf 3}-|4\rangle_{\bf 3}=0
\end{eqnarray}

Let us write down the structures coming from {\bf 15}, {\bf 20} and {\bf 84} of the $SU(4)$ operators.
\begin{eqnarray}
|1-{\bf 15}\rangle_{\bf (3,3)}&=&O^{\dagger i}_a O^a_i O^{\dagger j}_b O^c_k-{\rm traces}\nonumber\\
|2-{\bf 15}\rangle_{\bf (3,3)}&=&O^{\dagger j}_b O^a_i O^{\dagger i}_a O^c_k-{\rm traces}\nonumber\\
|6-{\bf 15}\rangle_{\bf (3,3)}&=&O^{\dagger j}_b O^c_k O^{\dagger i}_a O^a_i-{\rm traces}\nonumber\\
|7-{\bf 15}\rangle_{\bf (3,3)}&=&O^{\dagger i}_a O^c_k O^{\dagger j}_b O^a_i-{\rm traces}\nonumber\\
|3-{\bf 20}\rangle_{\bf (3,3)}&=&\Big(O^{\dagger [i}_{(b} O^{[a}_{(k} O^{\dagger l]}_{e)} O^{d]}_{m)}-{\rm traces}\Big)\epsilon_{ad}\epsilon^{ce}\epsilon_{il}\epsilon^{jm}\nonumber\\
|4-{\bf 84}\rangle_{\bf (3,3)}&=&\Big(O^{\dagger (j}_{(b} O^{[a}_{[i} O^{\dagger m)}_{e)} O^{d]}_{l]}-{\rm traces}\Big)\epsilon_{ad}\epsilon^{ce}\epsilon^{il}\epsilon_{km}\nonumber\\
|5-{\bf 84}\rangle_{\bf (3,3)}&=&\Big(O^{\dagger [i}_{[a} O^{(c}_{(k} O^{\dagger l]}_{d]} O^{e)}_{m)}-{\rm traces}\Big)\epsilon^{ad}\epsilon_{be}\epsilon_{il}\epsilon^{jm}
\end{eqnarray}
The first number is just an enumerating label, the second one gives the multiplet of $SU(4)$ to which it corresponds. In the case of the last three operators we let the redundant  antisymmetrizing brackets to make it more transparent.

The flavor and R-symmetry indices can be labeled by use of the structures in representation {\bf  3}. Let us adapt the following notation:
\begin{equation}
|1-{\bf 15}\rangle_{\bf (3,3)}=\Big(|1\rangle_{\bf 3},|1\rangle_{\bf 3}\Big)
\end{equation}
where the first $|1\rangle_{\bf 3}$ means that the R-symmetry indices correspond to the first structure in representation {\bf  3} of the list (\ref{sixstructures}) and the second $|1\rangle_{\bf 3}$ to the first structure of the corresponding list for the flavor indices.
\begin{eqnarray}\label{states33}
|1-{\bf 15}\rangle_{\bf (3,3)}&=&\Big(|1\rangle_{\bf 3},|1\rangle_{\bf 3}\Big) \ , \qquad\qquad
|2-{\bf 15}\rangle_{\bf (3,3)}=\Big(|2\rangle_{\bf 3},|2\rangle_{\bf 3}\Big) \ , \nonumber\\
|6-{\bf 15}\rangle_{\bf (3,3)}&=&\Big(|3\rangle_{\bf 3},|3\rangle_{\bf 3}\Big)\ , \qquad\qquad
|7-{\bf 15}\rangle_{\bf (3,3)}=\Big(|4\rangle_{\bf 3},|4\rangle_{\bf 3}\Big) \ , \nonumber\\
|3-{\bf 20}\rangle_{\bf (3,3)}&=&\Big(|5\rangle_{\bf 3},|6\rangle_{\bf 3}\Big)-\frac{1}{2}\Big(|1\rangle_{\bf 3},|1\rangle_{\bf 3}\Big)+\frac{1}{2}\Big(|2\rangle_{\bf 3},|2\rangle_{\bf 3}\Big)-\frac{1}{2}\Big(|3\rangle_{\bf 3},|3\rangle_{\bf 3}\Big)+\frac{1}{2}\Big(|4\rangle_{\bf 3},|4\rangle_{\bf 3}\Big) \ ,\nonumber\\
|4-{\bf 84}\rangle_{\bf (3,3)}&=&\Big(|5\rangle_{\bf 3},|5\rangle_{\bf 3}\Big)-\frac{1}{6}\Big(|1\rangle_{\bf 3},|1\rangle_{\bf 3}\Big)-\frac{1}{6}\Big(|2\rangle_{\bf 3},|2\rangle_{\bf 3}\Big)-\frac{1}{6}\Big(|3\rangle_{\bf 3},|3\rangle_{\bf 3}\Big)-\frac{1}{6}\Big(|4\rangle_{\bf 3},|4\rangle_{\bf 3}\Big) \ ,\nonumber\\
|5-{\bf 84}\rangle_{\bf (3,3)}&=&\Big(|6\rangle_{\bf 3},|6\rangle_{\bf 3}\Big)-\frac{1}{6}\Big(|1\rangle_{\bf 3},|1\rangle_{\bf 3}\Big)-\frac{1}{6}\Big(|2\rangle_{\bf 3},|2\rangle_{\bf 3}\Big)-\frac{1}{6}\Big(|3\rangle_{\bf 3},|3\rangle_{\bf 3}\Big)-\frac{1}{6}\Big(|4\rangle_{\bf 3},|4\rangle_{\bf 3}\Big) \ . \nonumber\\
\end{eqnarray}
We see that it is possible to write down 36 different structures, there are six different ways to put R-symmetry indices and 6 ways for the flavor indices. Since there are only three independent structures for one type of indices there are only 9 linear independent structures if we consider both types of the indices at the same time. In (\ref{states33}) we wrote down only 7 linear independent structures, 2 remaining ones correspond to {\bf 45} and ${\bf \overline{45}}$ of $SU(4)$ and don't correspond to any trace operators, that's why we are not considering them.

In general if we act with the Hamiltonian on these structures we will have structures which not immediately match with the structures in (\ref{states33}). Let us go give here the list of the relations which we used to obtain the mixing matrix in the main text.

We can immediately find the relations
\begin{equation}
{\rm Tr}\ \Big(|1\rangle_{\bf 3}+|3\rangle_{\bf 3},|5/6\rangle_{\bf 3}\Big)=0\ , \qquad
{\rm Tr}\ \Big(|2\rangle_{\bf 3}+|4\rangle_{\bf 3},|5/6\rangle_{\bf 3}\Big)=0 \ ,
\end{equation}
where we used 
\begin{equation}
{\rm Tr}\ \Big(|1\rangle_{\bf 3},|5/6\rangle_{\bf 3}\Big)=-{\rm Tr}\ \Big(|3\rangle_{\bf 3},|5/6\rangle_{\bf 3}\Big) \ , \qquad
{\rm Tr}\ \Big(|2\rangle_{\bf 3},|5/6\rangle_{\bf 3}\Big)=-{\rm Tr}\ \Big(|4\rangle_{\bf 3},|5/6\rangle_{\bf 3}\Big) \ .
\end{equation}

Other relations which we used are

\small
\begin{eqnarray}
{\rm Tr}\ \Big(|6\rangle_{\bf 3},|5\rangle_{\bf 3}\Big)&=&-{\rm Tr}\ \Big(|5\rangle_{\bf 3},|6\rangle_{\bf 3}\Big)+2\ {\rm Tr}\ \Big(|1\rangle_{\bf 3},|1\rangle_{\bf 3}\Big)-2\ {\rm Tr}\ \Big(|2\rangle_{\bf 3},|2\rangle_{\bf 3}\Big)\nonumber\\
&=&{\rm Tr} \ |1-{\bf 15}\rangle_{\bf (3,3)}-{\rm Tr} \ |2-{\bf 15}\rangle_{\bf (3,3)}-{\rm Tr} \ |3-{\bf 20}\rangle_{\bf (3,3)}
 \ ,
\nonumber
\end{eqnarray}
\begin{eqnarray}
{\rm Tr}\ \Big(|1\rangle_{\bf 3},|3\rangle_{\bf 3}\Big)&=&{\rm Tr}\ \Big(|3\rangle_{\bf 3},|1\rangle_{\bf 3}\Big)={\rm Tr}\ \Big(|2\rangle_{\bf 3},|2\rangle_{\bf 3}\Big)-\frac{1}{2}{\rm Tr}\ \Big(|5\rangle_{\bf 3},|5\rangle_{\bf 3}\Big)-\frac{1}{2}{\rm Tr}\ \Big(|6\rangle_{\bf 3},|6\rangle_{\bf 3}\Big)\nonumber\\
&=&-\frac{1}{3}{\rm Tr} \ |1-{\bf 15}\rangle_{\bf (3,3)}+\frac{2}{3}{\rm Tr} \ |2-{\bf 15}\rangle_{\bf (3,3)}-\frac{1}{2}{\rm Tr} \ |4-{\bf 84}\rangle_{\bf (3,3)}-\frac{1}{2}{\rm Tr} \ |5-{\bf 84}\rangle_{\bf (3,3)}
 \ ,
\nonumber\\
{\rm Tr}\ \Big(|2\rangle_{\bf 3},|4\rangle_{\bf 3}\Big)&=&{\rm Tr}\ \Big(|4\rangle_{\bf 3},|2\rangle_{\bf 3}\Big)={\rm Tr}\ \Big(|1\rangle_{\bf 3},|1\rangle_{\bf 3}\Big)-\frac{1}{2}{\rm Tr}\ \Big(|5\rangle_{\bf 3},|5\rangle_{\bf 3}\Big)-\frac{1}{2}{\rm Tr}\ \Big(|6\rangle_{\bf 3},|6\rangle_{\bf 3}\Big)\nonumber\\
&=&\frac{2}{3}{\rm Tr} \ |1-{\bf 15}\rangle_{\bf (3,3)}-\frac{1}{3}{\rm Tr} \ |2-{\bf 15}\rangle_{\bf (3,3)}-\frac{1}{2}{\rm Tr} \ |4-{\bf 84}\rangle_{\bf (3,3)}-\frac{1}{2}{\rm Tr} \ |5-{\bf 84}\rangle_{\bf (3,3)}\nonumber
 \ ,
\end{eqnarray}
\begin{eqnarray}
{\rm Tr}\ \Big(|1\rangle_{\bf 3},|2\rangle_{\bf 3}\Big)&=&{\rm Tr}\ \Big(|3\rangle_{\bf 3},|4\rangle_{\bf 3}\Big)={\rm Tr}\ \Big(|1\rangle_{\bf 3},|1\rangle_{\bf 3}\Big)-\frac{1}{2}{\rm Tr}\ \Big(|5\rangle_{\bf 3},|6\rangle_{\bf 3}\Big)-\frac{1}{2}{\rm Tr}\ \Big(|6\rangle_{\bf 3},|6\rangle_{\bf 3}\Big)\nonumber\\
&=&\frac{1}{3}{\rm Tr} \ |1-{\bf 15}\rangle_{\bf (3,3)}+\frac{1}{3}{\rm Tr} \ |2-{\bf 15}\rangle_{\bf (3,3)}-\frac{1}{2}{\rm Tr} \ |3-{\bf 20}\rangle_{\bf (3,3)}-\frac{1}{2}{\rm Tr} \ |5-{\bf 84}\rangle_{\bf (3,3)}
 \ ,
\nonumber\\
{\rm Tr}\ \Big(|1\rangle_{\bf 3},|4\rangle_{\bf 3}\Big)&=&{\rm Tr}\ \Big(|3\rangle_{\bf 3},|2\rangle_{\bf 3}\Big)={\rm Tr}\ \Big(|2\rangle_{\bf 3},|2\rangle_{\bf 3}\Big)-\frac{1}{2}{\rm Tr}\ \Big(|5\rangle_{\bf 3},|5\rangle_{\bf 3}\Big)+\frac{1}{2}{\rm Tr}\ \Big(|5\rangle_{\bf 3},|6\rangle_{\bf 3}\Big)\nonumber\\
&=&\frac{1}{3}{\rm Tr} \ |1-{\bf 15}\rangle_{\bf (3,3)}+\frac{1}{3}{\rm Tr} \ |2-{\bf 15}\rangle_{\bf (3,3)}+\frac{1}{2}{\rm Tr} \ |3-{\bf 20}\rangle_{\bf (3,3)}-\frac{1}{2}{\rm Tr} \ |4-{\bf 84}\rangle_{\bf (3,3)}\nonumber
 \ ,
\end{eqnarray}
\begin{eqnarray}
{\rm Tr}\ \Big(|2\rangle_{\bf 3},|1\rangle_{\bf 3}\Big)&=&{\rm Tr}\ \Big(|4\rangle_{\bf 3},|3\rangle_{\bf 3}\Big)={\rm Tr}\ \Big(|2\rangle_{\bf 3},|2\rangle_{\bf 3}\Big)+\frac{1}{2}{\rm Tr}\ \Big(|5\rangle_{\bf 3},|6\rangle_{\bf 3}\Big)-\frac{1}{2}{\rm Tr}\ \Big(|6\rangle_{\bf 3},|6\rangle_{\bf 3}\Big)\nonumber\\
&=&\frac{1}{3}{\rm Tr} \ |1-{\bf 15}\rangle_{\bf (3,3)}+\frac{1}{3}{\rm Tr} \ |2-{\bf 15}\rangle_{\bf (3,3)}+\frac{1}{2}{\rm Tr} \ |3-{\bf 20}\rangle_{\bf (3,3)}-\frac{1}{2}{\rm Tr} \ |5-{\bf 84}\rangle_{\bf (3,3)}
 \ ,
\nonumber\\
{\rm Tr}\ \Big(|2\rangle_{\bf 3},|3\rangle_{\bf 3}\Big)&=&{\rm Tr}\ \Big(|4\rangle_{\bf 3},|1\rangle_{\bf 3}\Big)={\rm Tr}\ \Big(|1\rangle_{\bf 3},|1\rangle_{\bf 3}\Big)-\frac{1}{2}{\rm Tr}\ \Big(|5\rangle_{\bf 3},|5\rangle_{\bf 3}\Big)-\frac{1}{2}{\rm Tr}\ \Big(|5\rangle_{\bf 3},|6\rangle_{\bf 3}\Big)\nonumber\\
&=&\frac{1}{3}{\rm Tr} \ |1-{\bf 15}\rangle_{\bf (3,3)}+\frac{1}{3}{\rm Tr} \ |2-{\bf 15}\rangle_{\bf (3,3)}-\frac{1}{2}{\rm Tr} \ |3-{\bf 20}\rangle_{\bf (3,3)}-\frac{1}{2}{\rm Tr} \ |4-{\bf 84}\rangle_{\bf (3,3)}
 \ . \nonumber\\ 
\end{eqnarray}
\normalsize
\section{Relations among the operator structures of (3,1) and (1,3) }
We can write down the singlet structures for the $SU(2)$ indices of the length four structures in two ways and the decomposition of $\bf 2\otimes 2\otimes 2 \otimes 2$ tells us that there only two singlets. It means that there are no linear relation among the singlet structures and we can consider them as the basis structures.

The singlets are 
\begin{eqnarray}\label{singlettstructures}
|1\rangle_{\bf 1} =O^\dagger_aO^aO^\dagger_b O^b \ , \qquad |2\rangle_{\bf 1} =O^\dagger_aO^bO^\dagger_b O^a \ .
\end{eqnarray}

Since there are 6 different ways to put the indices corresponding to the representation $\bf 3$, there are 12 structures which correspond to the structures of the type $(\bf 3,1)$ and 12 for $(\bf 1,3)$. 
Since the relations for $(\bf 3,1)$ or $(\bf 1,3)$ are similar we concentrate here only on the $(\bf 3,1)$ structures.

The four of the total six ({\bf 3,1})-structures come from {\bf 15} of ABJM. Let us see how they look like. 
Consider $|1\rangle_{\bf 15}+|1\rangle_{\bf 1}$ and replace $Y^A$ by $O^{\dagger i}_a$
\begin{eqnarray}
Y^C Y^\dagger_CY^B Y_A^\dagger&=&O^{\dagger i}_a O^a_i O^{\dagger j}_b O^c_k=\epsilon_{bd}\epsilon^{jl}O^{\dagger i}_a O^a_i O^{\dagger d}_l O^c_k\nonumber\\
&=&\epsilon_{bd}\epsilon^{jl}O^{\dagger i}_a O^a_i\Big(O^{\dagger (d}_{(l}O^{c)}_{k)}+O^{\dagger (d}_{[l}O^{c)}_{k]}+O^{\dagger [d}_{(l}O^{c]}_{k)}+O^{\dagger [d}_{[l}O^{c]}_{k]}\Big)
\end{eqnarray}
The second term corresponds to the structure which transforms in representation {\bf (3,1)}. Let us consider it.
\begin{eqnarray}
&&\frac{1}{4}\epsilon_{bd}\epsilon^{jl} O^{\dagger i}_a O^a_i\Big(O^{\dagger d}_{l}O^c_k+O^{\dagger c}_{l}O^d_k-O^{\dagger d}_{k}O^c_l-O^{\dagger c}_{k}O^d_l\big)\nonumber\\
&=&\frac{1}{4} O^{\dagger i}_aO^a_i\big(O^{\dagger j}_bO^{c}_k+\epsilon_{bd}\epsilon^{ce}O^{\dagger j}_eO^d_k-\epsilon^{jl}\epsilon_{km}O^{\dagger m}_b O^c_l-\epsilon_{bd}\epsilon^{jl}\epsilon^{ce}\epsilon_{km}O^{\dagger m}_e O^d_l\Big)\nonumber\\
&=&\frac{1}{4} O^{\dagger i}_aO^a_i\big(O^{\dagger j}_bO^{c}_k+\left(\delta^e_b\delta^c_d-\delta^c_b\delta^e_d\right)O^{\dagger j}_e O^d_k-\left(\delta_m^j\delta_k^l-\delta_k^j\delta_m^l\right)O^{\dagger m}_b O^c_l\nonumber\\
&&-\left(\delta_b^e\delta_d^c-\delta_b^c\delta_d^e\right)\left(\delta_m^j\delta_k^l-\delta^j_k\delta^l_m\right)O^{\dagger m}_eO^d_l\nonumber\\
&=&\frac{1}{2}\delta_k^j O^{\dagger i}_a O^a_i O^{\dagger m}_b O^c_m-\frac{1}{4}\delta_k^j\delta^c_b O^{\dagger i}_a O^a_i O^{\dagger m}_d O^d_m
\end{eqnarray}
Therefore, the four ({\bf 3,1})-structures descending form the four {\bf 15}-structures are
\begin{eqnarray}
|1\rangle_{\bf (3,1)}&=&O^{\dagger i}_a O^a_i O^{\dagger m}_b O^c_m-{\rm trace} \ ,\qquad\qquad
|2\rangle_{\bf (3,1)}=O^{\dagger m}_b O^a_i O^{\dagger i}_a O^c_m-{\rm trace} \ ,\nonumber\\
|3\rangle_{\bf (3,1)}&=&O^{\dagger m}_b O^c_m O^{\dagger i}_a O^a_i-{\rm trace} \ , \qquad\qquad
|4\rangle_{\bf (3,1)}=O^{\dagger i}_a O^b_m O^{\dagger m}_c O^a_i-{\rm trace} \ .
\end{eqnarray}
The remaining two structures come from the ABJM multiplet transforming under 45 and $\overline 45$ of ABJM and do not correspond to any trace invariant operators.

To find the necesarry relations among the 12 different structures of ({\bf 3,1}) we use the same trick as in the previous section of the appendix. We write 
\begin{eqnarray}
|1\rangle_{\bf (3,1)}&=&\Big(|1\rangle_{\bf 3},|1\rangle_{\bf 1}\Big)\ ,\qquad\qquad 
|2\rangle_{\bf (3,1)}=\Big(|2\rangle_{\bf 3},|2\rangle_{\bf 1}\Big) \ ,\nonumber\\
|3\rangle_{\bf (3,1)}&=&\Big(|3\rangle_{\bf 3},|1\rangle_{\bf 1}\Big)\ ,\qquad\qquad
|4\rangle_{\bf (3,1)}=\Big(|4\rangle_{\bf 3},|2\rangle_{\bf 1}\Big) \ .
\end{eqnarray}

If we apply the Hamiltonian to these structures we find structures which do not match with the above structures. The structures which we need to identify are
\begin{eqnarray}
O^{\dagger i}_a O^c_i O^{\dagger j}_b O^a_j+O^{\dagger i}_b O^a_i O^{\dagger j}_a O^c_j=\Big(|2+4\rangle_{\bf 3},|1\rangle_{\bf 1}\Big) \ ,\nonumber\\
O^{\dagger i}_a O^a_j O^{\dagger j}_b O^c_i+O^{\dagger i}_b O^c_j O^{\dagger j}_a O^a_i=\Big(|1+3\rangle_{\bf 3},|1\rangle_{\bf 1}\Big) \ .
\end{eqnarray}
By use of (\ref{rel}) we find
\begin{eqnarray}
O^{\dagger i}_a O^c_i O^{\dagger j}_b O^a_j+O^{\dagger i}_b O^a_i O^{\dagger j}_a O^c_j=\Big(|2+4\rangle_{\bf 3},|1\rangle_{\bf 1}\Big)=\Big(|1+3\rangle_{\bf 3},|1\rangle_{\bf 1}\Big)=|1\rangle_{\bf (3,1)}+|3\rangle_{\bf (3,1)} \ ,\nonumber\\
O^{\dagger i}_a O^a_j O^{\dagger j}_b O^c_i+O^{\dagger i}_b O^c_j O^{\dagger j}_a O^a_i=\Big(|1+3\rangle_{\bf 3},|2\rangle_{\bf 1}\Big)=\Big(|2+4\rangle_{\bf 3},|2\rangle_{\bf 1}\Big)=|2\rangle_{\bf (3,1)}+|4\rangle_{\bf (3,1)} \ .
\end{eqnarray}

\section{The general form of the mixing operator from the six-vertex diagram}\label{gaugechoice}

%
%
%
%
The terms in the equation (\ref{ansatz}) are not linearly independent and in the main text of this article we have chosen a specific choice of the $a_n$ which allowed to eliminate the linear dependencies and  write the potential only with six terms. It is not necessary to make a concrete choice. Actually, to obtain the mixing operator of the main text we first computed the mixing operator starting from the full ansatz of the potential  (\ref{ansatz}) and only then inserted the coefficients $a_n$ from the solution of $V^{bos}_{a_n}=V^{bos}$. We put this formulae into the appendix since it allows the reader to write down the mixing operator in a different form then in eq.  ({\ref{hful}}). Which terms in the ansatz $V^{bos}_{a_n}$ are allowed to be set to zero can be decided looking at the linear relations (\ref{depend}).

The mixing operator from the six-vertex diagram derived from the ansatz in eq. (\ref{ansatz}) is
%
\begin{eqnarray}
\Gamma^{V\otimes\bar  V\otimes V}&=&\frac{N^2}{16 \pi^2}\Big( K^{\bar V\otimes V}\left(3a_1\hat K^{\bar V\otimes V}+3a_4\hat K^{V\otimes \bar V}+a_2 \left(\hat 1+\hat K\hat P+\hat P\hat K\right)+3a_3 \hat P\right)\nonumber\\
&+& K^{V\otimes \bar V}\left(3a_5\hat K^{\bar V\otimes V}+3a_8\hat K^{V\otimes \bar V}+a_6 \left(\hat 1+\hat K\hat P+\hat P\hat K\right)+3a_7 \hat P\right)\nonumber\\
&+&KP\left(a_9\hat K^{\bar V\otimes V}+a_{12}\hat K^{V\otimes \bar V}+a_{11}\hat P+a_{10}\hat K\hat P+a_{17}\hat P\hat K+a_{18}\hat 1\right)\nonumber\\
&+&PK\left(a_9\hat K^{\bar V\otimes V}+a_{12}\hat K^{V\otimes \bar V}+a_{11}\hat P+a_{10}\hat P\hat K+a_{17}\hat 1+a_{18}\hat K\hat P\right)\nonumber\\
&+&1\left(a_9\hat K^{\bar V\otimes V}+a_{12}\hat K^{V\otimes \bar V}+a_{11}\hat P+a_{10}\hat 1+a_{17}\hat K\hat P+a_{18}\hat P\hat K\right)\nonumber\\
&+& P\left(3a_{13}\hat K^{\bar V\otimes V}+3a_{16}\hat K^{V\otimes \bar V}+a_{14}\left(\hat 1+\hat K\hat P+\hat P\hat K\right)+3a_{15}\hat P\right)\Big) \ .
\end{eqnarray}
To obtain  the $\Gamma^{V\otimes\bar  V\otimes V}$ piece of the dilatation operator one needs to exchange $K^{\bar V\otimes V}$ by $K^{V\otimes \bar V}$ and additionally the following coefficients
\begin{eqnarray}
a_1 &\leftrightarrow & a_8 \ , \qquad a_2 \leftrightarrow  a_6 \ , \qquad \ \  a_3 \leftrightarrow  a_7 \ ,\nonumber \\ 
a_4 &\leftrightarrow & a_5\ , \qquad a_9 \leftrightarrow  a_{12}\ , \qquad a_{13} \leftrightarrow a_{16} \ .
\end{eqnarray} 
Formally, $\Gamma^{V\otimes\bar  V\otimes V}$ looks the same, but $PK$ and $KP$ act now on the $\bar V \otimes V \otimes\bar V$ spaces.

\section{Representations of length four structures }
A general length four operator transforming under ${\bf (2,2)}^4$ of $SU(2)_R \times SU(2)$ will decompose into the irreducible representations as follows
\begin{eqnarray} \label{decompose}
{\bf (2,2)}^4= {\bf (1,1)}^4\oplus {\bf (1,3)}^6 \oplus   {\bf (1,5)}^2 \oplus  {\bf (3,1)}^6 \oplus   {\bf (3,3)}^9 \oplus  {\bf (3,5)}^3 \oplus  {\bf (5,1)}^2 \oplus {\bf (5,3)}^3\oplus  {\bf (5,5)}\nonumber\\
\end{eqnarray}
From \cite{Minahan:2008hf} we know that in the case of the length four operators in ABJM the integrability manifests in the degeneracy of the the trace invariant operators transforming in the representation $\bf 15$ of $SU(4)$.  In the main text of this article we checked if the degeneracy still holds among the operators $\bf (3,1)$, $\bf (1,3)$ and $\bf (3,3)$ which descend from those in the {\bf 15} of ABJM in the notation of \cite{Minahan:2008hf}.

From (\ref{decompose}) we see that there are actually more than 4 structures in each of the representations $\bf (3,3)$, $\bf (1,3)$ and $\bf (3,1)$. This comes from the fact that some of them are also present in other multiplets. The decomposition of all length four operators of ABJM under $SU(2)_R \times SU(2)$ goes as follows
\begin{eqnarray}\label{decomposition}
&&{\bf 1} \rightarrow {\bf (1,1)}\nonumber\\
&&{\bf 20} \rightarrow {\bf (1,1)} + {\bf (3,3)} + {\bf (1,5)} + {\bf (5,1)}\nonumber\\
&&{\bf 15} \rightarrow {\bf (1,3)} + {\bf (3,1)} + {\bf (3,3)}\nonumber\\
&&{\bf 45} \rightarrow {\bf (1,3)} + {\bf (3,1)} + {\bf (3,3)} + {\bf (5,3)}+ {\bf (3,5)}\nonumber\\
&&{\bf \overline{45}} \rightarrow {\bf (1,3)} + {\bf (3,1)} + {\bf (3,3)} + {\bf (5,3)}+ {\bf (3,5)}\nonumber\\
&&{\bf 84} \rightarrow {\bf (1,1)} +  {\bf (3,3)}^2 + {\bf (1,5)} + {\bf (5,1)}+ {\bf (3,5)} + {\bf (5,3)} + {\bf (5,5)}
\end{eqnarray}

The structures coming from the $\bf 45$ and the $\bf \bar{45}$ do not correspond to any trace invariant operators because they get a minus under cyclic permutation.

\end{document}